\documentclass[aps,prl,reprint,nofootinbib]{revtex4-1}

\usepackage{amsmath,amssymb,amsthm}
\usepackage{bbm}
\usepackage{graphicx}
\usepackage{hyperref}
\theoremstyle{definition}

\theoremstyle{remark}
\newtheorem*{rem*}{Remark}
\newtheorem*{note*}{Note}

\usepackage{xcolor}

\begin{document}
\title{Attractors, Geodesics, and the Geometry of Moduli Spaces}
\author{Fabian Ruehle}
\affiliation{Department of Physics and Department of Mathematics, Northeastern University, Boston, MA 02115, USA}
\affiliation{NSF Institute for Artificial Intelligence and Fundamental Interactions, Boston, MA, USA}
\email{f.ruehle@northeastern.edu}
\author{Benjamin Sung}
\affiliation{Department of Mathematics, University of California, Santa Barbara, CA 93106, USA}
\email{bsung@ucsb.edu}

\begin{abstract}
We connect recent conjectures and observations pertaining to geodesics, attractor flows, Laplacian eigenvalues and the geometry of moduli spaces by using that attractor flows are geodesics. For toroidal compactifications, attractor points are related to (degenerate) masses of the Laplacian on the target space, and also to the Laplacian on the moduli space. We also explore compactifications of M-Theory to $5$D on a Calabi-Yau threefold and argue that geodesics are unique in a special set of classes, providing further evidence for a recent conjecture by Raman and Vafa. Finally, we describe the role of the marked moduli space in $4$d $\mathcal{N} = 2$ compactifications. We study split attractor flows in an explicit example of the one-parameter family of quintics and discuss setups where flops to isomorphic Calabi-Yau manifolds exist.
\end{abstract}

\maketitle

\section{Introduction}
In the context of the swampland program, the structure of moduli spaces of Calabi-Yau manifolds, as well as its relation to particle masses, has been studied extensively. In this paper, we aim to elucidate connections between recent observations and conjectures.

In~\cite{Ahmed:2023cnw}, Ahmed and one of the authors proved a connection between attractor points and degeneracy of eigenvalues of the scalar Laplacian for toroidal compactifications. It was also conjectured (and some numerical evidence was presented), that this relation extends to Calabi-Yau manifolds more broadly.

In~\cite{Etheredge:2023zjk}, the authors established that the solution of a gradient flow with constant length gradients are geodesics. They moreover showed that if the Laplacian, i.e., the divergence of the gradient, is also constant, then the function solves a Laplace equation.

In~\cite{Raman:2024fcv}, the authors introduce the marked moduli space and a conjecture which states in its strong version that the geodesic between any two points in the marked moduli space is unique.

Our main contribution is to point out the correspondence between geodesics and attractor flows, thereby connecting the above three recent conjectures. To our knowledge, this correspondence has not been explored in the context of these developments in the swampland program; thus, our goal is to make these connections precise, and to study the basic implications in examples.

This article is organized as follows. We begin by reviewing the fact that attractor flows are geodesics, and the connection with the gradient flows studied in~\cite{Etheredge:2023zjk}. We then discuss attractor flows in Calabi-Yau moduli spaces and the uniqueness of flows associated with a single charge. We further describe the significance of the marked moduli space described in~\cite{Raman:2024fcv} by finding an infinite number of geodesics between two points realized via flop transitions in the unmarked space. Focusing on the case of abelian varieties, we explicitly show that the attractor flow for a class of $1/2$ BPS states is a geodesic and is simultaneously an eigenfunction of the moduli space Laplacian, and we show that this coincides with eigenvalues of the Laplacian on tori, as explored in~\cite{Ahmed:2023cnw}. Finally, we study $4$d $\mathcal{N}=2$ moduli spaces which feature split attractor flows, and we describe the resulting geodesics.

\section{Relation between geodesics and attractor flows}
In the following, we will briefly review the attractor flow mechanism following~\cite{Ferrara:1995ih,Ferrara:1997tw,Moore:1998pn,Denef:2000nb,Denef:2001xn}. The metric of a static, spherically symmetric BPS object at the origin of space is
\begin{align}
    ds^2=-e^{2U}dt^2+e^{-2U}(dx^i)^2\,,
\end{align}
where $U=U(r^{-1})$ is a real function of the radial distance $r=|\vec{x}|$ and $U(r^{-1}=0)=0$ for an asymptotically flat spacetime. The resulting BPS equations can be written as gradient flow equations~\cite{Ferrara:1997tw,Moore:1998pn}
\begin{align}
    \label{eq:AttractorEqns}
    \dot U=-e^{U}|Z|\,,\quad
    \dot z^a=-2e^{U}g^{a\overline{b}}\overline\partial_{\overline{b}}|Z|\,,
\end{align}
where the dot denotes derivatives w.r.t.~$\tau=r^{-1}$, $z^a$ are moduli fields, $g_{a\overline{b}}=\partial_a\overline{\partial}_{\overline{b}}K$ is the Weil-Petersson (WP) metric on moduli space, and $Z$ is the central charge. The attractor flow equation can also be obtained from a Polyakov action for a static string in $(U,z)$ space with action~\cite{Denef:2001xn}
\begin{align}
    \label{eq:PolyakovAction}
    S=-\int dt\, d\tau\; \left[e^{2U}V(\dot U^2+||\dot z||^2)\right]^{-1/2}\,,
\end{align}
where 
\begin{align}
V=|Z|^2+2\frac{D-2}{D-3}g^{a\overline{b}}\partial_a|Z|\overline{\partial}_{\overline{b}}|Z| 
\end{align}
is the black hole potential in $D$ dimensions. 
% With this, the attractor flow trajectory in moduli space can be written as a gradient flow~\cite{Denef:2001xn, Perz:2008kh},
% \begin{align}
%     \label{eq:GradientFlow}
%     \dot{z}^a=-g^{a\overline{b}}\overline\partial_{\overline{b}}|Z|\,.
% \end{align}
Equation~\eqref{eq:PolyakovAction} relates attractor flows to geodesics. One of the endpoints of the string is fixed at spatial infinity $\tau=0$. Extremizing the action fixes the other endpoint at an attractor point and the string tension is minimized if the string is a geodesic. This is useful since geodesics are second-order PDE while the attractor flow is a first-order PDE. Another way of seeing the relation between attractor flows and geodesics uses Theorem in~\cite[Equation 2]{Etheredge:2023zjk}: This theorem states that if the RHS of~\eqref{eq:AttractorEqns} has constant length with respect to the WP metric (which it does for attractors), then $z^a$ satisfies the geodesic equation.

\section{Uniqueness of geodesics in CY moduli spaces}
Let us next discuss the question of uniqueness of geodesics in the marked moduli space. From the point of view of attractor flows, the flows (and hence geodesics) are unique in the moduli space away from singular points where $Z=0$. To see this, we use that $|Z|$ is a strictly decreasing function along the flow. Then, since the trajectory through moduli space~\eqref{eq:AttractorEqns} follows a gradient flow whose potential is a strictly decreasing function, the flow will end at a unique minimum, which is the attractor point.

To see that $|Z|$ is strictly decreasing (excluding the $|Z|=0$ case, to which we will return below), note that the first equation in~\eqref{eq:AttractorEqns} implies that $U$ is a strictly decreasing function of $\tau$. One can show that along the attractor flow, the central charge evolves as
\begin{align}
\label{eq:CentralchargeFlow}
    |\dot Z|=-4e^{U}g^{a\overline{b}}\partial_a|Z|\;\overline{\partial}_{\overline{b}}|Z|:=-4e^U(\nabla|Z|)^2\,,
\end{align}
which is then also strictly decreasing along the flow. 

The assumption that $U$ decreases along the flow is clearly invalidated if $|Z|=0$. In such a situation, the question of the existence of multiple basins of attractions for the attractor flow was already raised in Moore's paper~\cite[Section 9.2]{Moore:1998pn}, and Denef proposed a resolution in~\cite{Denef:2000nb,Denef:2001xn}. Essentially, different basins of attractions are linked to split attractor flows, where the flow line of a particle $\Gamma$ crosses a wall of marginal stability. At the wall, it decays into two particles, $\Gamma\to\Gamma_1+\Gamma_2$, with $Z(\Gamma)=Z(\Gamma_1)+Z(\Gamma_2)$. Each $\Gamma_i$ then follows its own gradient flow to its respective attractor points, and the corresponding geodesics become that of a multi-pronged string. The particle with $Z=0$ ($\Gamma_1$, say), will end at a singularity and $\Gamma_2$ will continue its flow. Since $\Gamma_2=\Gamma-\Gamma_1$, its attractor point will not be the original end of the flow for $\Gamma$, but will instead be another point which coincides with the attractor point of $\Gamma$ when starting in the other basin of attraction on the other side of the singularity where the flow of $\Gamma_1$ ended.
%\BS{So $\Gamma$ isn't the particle with $Z = 0$, but $\Gamma_1$ is? Is there perhaps a figure we can add to show this?} \fr{We could reproduce Figure 5 from~\cite{Denef:2001xn}, but maybe we do not need that level of detail, since it is not our work?}

While our above discussion illustrates that the flow associated to a charge $\Gamma$ is unique, it does not imply the strong version of the conjecture in~\cite{Raman:2024fcv} that the geodesics between any two points in the marked moduli space are unique. Indeed, consider a situation where the flow lines of two particles $\Gamma_1$ and $\Gamma_2$ intersect more than once in the marked moduli space, say at two points $P$ and $Q$. Since the attractor flows of each $\Gamma_i$ are geodesics, this would mean that there are two geodesics connecting $P$ and $Q$. While we do not have a proof that this cannot happen, the hope is that studying the attractor flow equation, which is a first order PDE given by gradient flow along a potential with constant length, is simpler than showing uniqueness of the second order PDE that underlies the geodesics equation. Since attractor points are dense in the moduli space~\cite{Lam:2020qge}, and their flows are unique, the corresponding geodesics are unique on a dense set of points. Smoothness of the moduli space metric then suggests that geodesics are unique. Note that this statement breaks down at the boundary of moduli space (at singularities).

Let us also discuss the uniqueness of geodesics from the perspective of the K\"ahler moduli space of the mirror-dual. In~\cite{Brodie:2021nit}, geodesics in the extended K\"ahler cone of all Calabi-Yau manifolds with Picard rank 2 were studied. Quantum corrections were neglected, such that the expressions can for example be applied to study the vector multiplet moduli space of M-Theory compactifications to 5D $\mathcal{N}=1$ theories. The moduli space of such manifolds can be grouped into three classes which are determined by the pre-potential, and the geodesics equation can be solved uniquely and analytically for all cases. 

The behavior of the geodesics depends on the kind of primitive contraction encountered at the K\"ahler cone walls. At walls where a curve can be flopped, and the birational CY after the flop transition is isomorphic to the original CY, the geodesic gets ``reflected'' back onto itself. Hence, there is more than one geodesic (albeit only one shortest one) that connect two points; indeed, in the case where both K\"ahler cone walls support flops to isomorphic CYs, there are infinitely many such geodesics. 

The geodesics become unique once the marked moduli space is considered: Instead of reflecting the geodesic back at the K\"ahler cone wall, it passes the wall and enters the adjacent K\"ahler cone of the isomorphic CY. Each ``bounce'' off a flop wall, or equivalently each crossing into an adjacent cone in the extended K\"ahler cone of the Calabi-Yau, is associated with an element of the Coxeter group that describes the symmetry of the K\"ahler moduli space~\cite{Lukas:2022crp}. Hence the Coxeter group can be used to distinguish the different geodesics that connect two points. We should mention that the Coxeter symmetry is also a symmetry of the quantum corrections to the K\"ahler potential and hence should be applicable more generally. %\BS{Is there a clear proof of uniqueness for Picard rank 2?}\fr{I'd say we've done this. What we showed in our papers is that in this case, any CY can be classified into three cases depending on its intersection form. For each case, there is a unique solution to the geodesics equation.}\BS{should we add a sentence or two mentioning that uniqueness of geodesics can be shown by a direct construction in these cases?}

Finally, let us mention that the contractibility of the marked moduli space for homogeneous K\"ahler manifolds (i.e., for the scalar moduli spaces of $\mathcal{N}>2$ supergravity theories) is equivalent to the existence of a globally defined K\"ahler potential on this space~\cite{Loi:2015aaa}. %\BS{Unfortunately, they have an extra assumption that the manifold must be homogeneous as well. This is only true for the moduli space of $\mathcal{N} >2$ supergravity theories.}\fr{I added this caveat but would like to learn more about it when we chat.}
The latter has been conjectured in~\cite{Gomis:2015yaa} based on anomaly considerations and checked in multiple instances for resolutions of toroidal orbifolds in~\cite{Donagi:2017mhd}. The simplest examples of homogeneous spaces consist of symmetric spaces of the form $G/H$ for $G$ a connected Lie group. Although generic moduli spaces of $4$d $\mathcal{N} = 2$ supergravity theories are not of this form, they admit a period mapping to a homogeneous space, and it would be interesting to explore the implications of a global K\"ahler potential in these cases.

\section{Attractors and uniqueness of geodesics in 5D M-Theory}
Let us consider a 5D $\mathcal{N}=1$ theory obtained from compactifying M-Theory on a Calabi-Yau threefold $X$. In this setting, the vector multiplet moduli $z^a$ of the attractor equations~\eqref{eq:AttractorEqns} are mapped to the (reduced) K\"ahler moduli space of the Calabi-Yau threefold. The theory is mostly governed by the prepotential\footnote{Alternatively, we can consider Type IIA string theory compactified on the same CY to 4D $\mathcal{N}=2$ and discard quantum corrections.}
\begin{align}
    \mathcal{F}=\kappa_{abc}t^at^bt^c\,,
\end{align}
where $t^a$, $a=1,\ldots, h^{1,1}(X)$ are the K\"ahler parameters of~$X$. The vector multiplet moduli space is parameterized by the slice $\mathcal{F}=6$, introducing a relation among the $t^a$. Put differently, the overall volume corresponds to a hypermultiplet. In particular, as was worked out in~\cite{Brodie:2021ain, Brodie:2021nit}, the geodesics equation (with the hypermultiplet moduli kept constant) becomes 
\begin{align}
\label{eq:Geodesics}
    g_{ab}\dot z^a \dot z^b=E\,,
\end{align}
where $E\geq 0$ and $z^a=t^a/\sqrt[3]{\text{vol}(X)}$ are the K\"ahler parameter with the overall volume scaled out, such that they parameterize the vector multiplet moduli space. Equation~\eqref{eq:Geodesics} was used in~\cite{Brodie:2021ain,Brodie:2021nit} to explicitly construct analytic expressions for the solution to the geodesics equations in the vector multiplet moduli space for $h^{1,1}=2$, which shows by construction that the geodesics are unique in this case. Using~\eqref{eq:AttractorEqns} and \eqref{eq:CentralchargeFlow} (and transitioning to the complexified K\"ahler parameters), the geodesic equation~\eqref{eq:Geodesics} can be written in terms of the length of the gradient,
\begin{align}
    E=4 e^{2U}(\nabla|Z|)^2=-e^{U}\partial_\tau|Z|=\ddot U -\dot U^2\,.
\end{align}
Note that since the RHS is positive, the curvature of $U$ has to be positive and larger than $\dot U^2$.

Let us study BPS objects obtained from wrapping $M2$ branes on 2-cycles $\mathcal{C}$ or $M5$ branes on 4-cycles in this theory. For the latter, the 4-cycles are Poincare-dual to two-cycles, which in turn are mirror-dual to three-cycles $\Gamma$ in the mirror CY $\widetilde{X}$. The central charges of these objects are
\begin{align}
    Z_{M2}=\int_\mathcal{C} J\,,\quad Z_{M5}=\int_\Gamma \widetilde{\Omega}\,,
\end{align}
where $J=t^aD_a$ is the K\"ahler form on $X$ and $\widetilde{\Omega}$ is the holomorphic volume form on the mirror $\tilde{X}$. The periods $\widetilde{\Pi}_i=\int_{\Gamma_i}\widetilde{\Omega}=\partial_i \mathcal{F}$ can thus be written in terms of the prepotential of the original CY $X$. It is customary to normalize the central charge as
\begin{align}
    Z=\frac{\int_\Gamma \widetilde{\Omega}}{\sqrt{\int \widetilde{\Omega}\wedge\overline{\widetilde{\Omega}}}}=\frac{\widetilde{\Pi}\cdot\Gamma}{\sqrt{i\;\widetilde{\Pi}^\dagger\cdot\chi^{-1}\cdot\widetilde{\Pi}}}\,,
\end{align}
where $\chi$ is the intersection form on the middle homology. 

Since $\kappa_{abc}\geq0$ for K\"ahler cone generators (which are effective divisors) and $z^i\geq 0$ in the K\"ahler cone, we see again that the potential of the gradient flow is either convex or concave (but not both), i.e., the central charge is decreasing along the flow and the flow is unique. Together with our discussion above, this means that the geodesics are unique as well.

% In the previous paragraphs, we argued that flow lines of a single defect are unique. Since attractor points are dense in the moduli space~\cite{Lam:2020qge}, and their flows are unique (if they exist, i.e., at least in all cases where $|Z|\neq0$ in the interior of the moduli space), geodesics are unique on a dense set of points. Smoothness of the moduli space metric then suggests that geodesics are unique everywhere. Note that this statement breaks down at the boundary of moduli space (at singularities).\fr{I reworded this}

%If it were true that every geodesic arises as the gradient flow of some defect, and that two flow lines associated with two different defects never intersect more than twice, this would give a proof that geodesics in the marked moduli space are unique. 

For the case of the upper half plane, the group SL$(2,\mathbb{R})$ acts naturally on the flow lines associated with a single entropy functional, and hence relates flow lines between different defects. This could potentially be used to re-prove the uniqueness of geodesics, and we leave this avenue of investigation for future work.

%\TODO{Ben, can you put in what you wrote on Slack about this point}. Moreover, two flow lines associated with two different defects never intersect twice (which is also true for the upper half space). Together, this means that the geodesics in the marked moduli space are unique and hence, it is contractible.

\section{Abelian varieties}
Here we discuss the attractor flows, geodesics, and Laplacian modes for Abelian varieties. Let us start by discussing the spectrum of the scalar Laplacian on $T^{2n}=\mathbbm{C}^n/\Lambda$ where $\Lambda=\mathbbm{Z}^n\oplus\tau\mathbbm{Z}^n$. We will find it more convenient to work over the real numbers. For this we let the $n\times n$ matrix $\tau=X+iY$ and define the $2n\times 2n$ period matrix
\begin{align}
\label{eq:PeriodMatrix}
\Pi=\begin{pmatrix}
\mathbbm{1}&X\\0&Y
\end{pmatrix}\,.
\end{align}
The Laplacian eigenspectrum is in one-to-one correspondence with vectors in the dual lattice
\begin{align}
\check{\Lambda}=\{c\in\mathbbm{R}^{2n}~|~\langle c,v_i\rangle=2\pi k_i\}\,,
\end{align}
where $v_i$ are the column vectors of $\Pi$ and $k_i\in\mathbbm{Z}$. We can write any $c\in\check{\Lambda}$ as
\begin{align}
\label{eq:DualVector}
c&=2\pi(\Pi^{-1})^T\cdot\vec{k}=\frac{2\pi}{\text{det}(\Pi)}\text{Cof}(\Pi)\cdot\vec{k}
\end{align}
where $\text{Cof}(\Pi)=\det(\Pi)(\Pi^{-1})^T$ is the cofactor matrix. Note that $\det(\Pi)=\det(Y)\neq0$ for the torus to be non-degenerate. The eigenfunctions $\phi(x)$ for $x\in\mathbbm{R}^{2n}$ and eigenvalues $E_{\vec{k}}$ are then given by
\begin{align}
\label{eq:eigvalsTn}
\begin{split}
\phi_{\vec{c}}(x)&=e^{i\vec{c}\cdot\vec{x}}\,,\\
E_{\vec{k}}&=|c|^2=\frac{4\pi^2}{[\det(Y)]^2}\;\vec{k}^T\cdot\text{Cof}(\Pi)^T\text{Cof}(\Pi)\cdot\vec{k}\,.
\end{split}
\end{align}
We also define the volume-normalized eigenvalues 
\begin{align}
    e_{\vec{k}}=\text{vol}(T^{2n})^{\frac{1}{n}} E_{\vec{k}}=\left|\det \Pi\right| \;E_{\vec{k}}
\end{align}
For Abelian varieties, we are precisely in a situation where $(\nabla F)^2$ (with $F=\log|Z|$) is constant for $1/2$-BPS states~\cite{Etheredge_2022,Etheredge:2023usk}. We give a complementary, geometric argument for this statement, which also demonstrates that $\Delta F=\nabla_a(\nabla^a F)$ is constant. This means that $Z$ is an eigenfunction of the (moduli space) Laplacian, see~\cite[Equation 20]{Etheredge:2023zjk}.

More precisely, we will verify the following two conditions in the case that $Z$ is the central charge of a state which lies in $\mathrm{Sp}(n)$ orbit of the $D0$-brane:
\begin{enumerate}
\item 
$(\nabla F)^2$ is constant.
\item 
$\nabla_a(\nabla^a F)$ is constant.
\end{enumerate}
We note that every $1/2$-BPS object on $T^2$ in particular, satisfies this property.

The moduli space of $\tau$ is parametrized by the Siegel upper half space of complex symmetric matrices 
\begin{equation}\label{eqn:siegel}
Z = X + iY
\end{equation}
where $Y$ is positive-definite with metric
\begin{equation}\label{eqn:metric}
ds^2 = \mathrm{Tr}(Y^{-1} dZ Y^{-1} d \overline{Z}).
\end{equation}
Moreover this space is K\"ahler, and the metric can be written in the form
\[
g_{i\bar{j}} = \partial_i \partial_{\bar{j}}K(z, \bar{z})
\]
for some K\"ahler potential $K(z,\bar{z})$. The choice
\begin{align}\label{eqn:vdefn}
    K=-\log \mathcal{V}, \quad \mathcal{V} = \mathrm{det}(\Pi)
\end{align}
will give rise to the Weil-Petersson metric.

To prove the first claim, we can write the norm of the gradient as
\begin{align}
\label{eqn:euler}
(\nabla F)^2 = g^{i\bar{j}} \partial_i \mathrm{log}\left[\frac{Z_\tau(\mathcal{E})}{\sqrt{\mathcal{V}(z,\bar{z})}}\right] \partial_{\bar{j}} \mathrm{log}\left[\frac{Z_{\tau}(\mathcal{E})}{\sqrt{\mathcal{V}(z,\bar{z})}}\right]
\end{align}
where $Z_{\tau}(\mathcal{E})$ is the central charge of a bound state of $D(2p)$-branes, and the denominator is the K\"ahler potential. This quantity transforms as a scalar under an Sp$(n)$ action, and by assumption that $\mathcal{F}$ is in the orbit of the $D0$-brane, there exists $\Phi \in \text{Sp}(n)$ such that $\Phi \cdot \mathcal{F} = (0, \ldots , 1)$. Thus, it suffices to prove that Equation~\eqref{eqn:euler} is constant for a $D0$-brane state. In this case, we claim that
\begin{align*}
(\nabla F )^2 &= g^{i\bar{j}} \partial_{i} \mathrm{log} \frac{1}{\sqrt{\mathcal{V}(z,\bar{z})}} \partial_{\bar{j}} \mathrm{log} \frac{1}{\sqrt{\mathcal{V}(z,\bar{z})}} \\
& \propto g^{i\bar{j}} \partial_{i} \mathrm{log}(\mathcal{V}(z,\bar{z})) \partial_{\bar{j}} \mathrm{log} (\mathcal{V}(z,\bar{z}))
\end{align*}
must be constant. 

To see this, we note that a direct calculation implies:
\begin{align*}
g_{i\bar{j}} &= \partial_i \partial_{\bar{j}} \mathrm{log}(\mathcal{V}) \\&= -\frac{1}{\mathcal{V}^2}(\partial_i \mathcal{V} \partial_{\bar{j}} \mathcal{V}) + \frac{1}{\mathcal{V}}\partial_i \partial_{\bar{j}} \mathcal{V} \\
&= \partial_i \mathrm{log}(\mathcal{V}) \partial_{\bar{j}} \mathrm{log}(\mathcal{V}) + \frac{1}{\mathcal{V}} \partial_i \partial_{\bar{j}}\mathcal{V}
\end{align*}

Thus, we have
\[
(\nabla F)^2 \propto g^{i\bar{j}}(g_{i\bar{j}} - \frac{1}{\mathcal{V}} \partial_i \partial_{\bar{j}}\mathcal{V})
\]
and the first term is clearly constant. By~(\ref{eqn:vdefn}) and the fact that $\mathcal{V}$ is a function of only $z- \bar{z}$, the constancy of the second term is equivalent to the following linear algebraic statement: Let $M$ be a an $n \times n$ symmetric, positive definite matrix with ${n+1 \choose 2}$ variables in the upper triangular entries and let $H(f)$ denote the Hessian of a function $f$. Then 
\begin{align}
\label{eq:TrIdentity}
\frac{1}{\mathrm{det}(M)}\text{tr}[H^{-1}(\log\det M)H(\det M)]={n \choose 2}
\end{align}
is a constant that depends on $n$ but not on $M$.

%Indeed, it suffices to show the vanishings:
%\begin{align*}
%\partial_a \left(g^{i\bar{j}} \partial_i \mathrm{log}(K(z,\bar{z}))\; \partial_{\bar{j}} \mathrm{log}(K(z,\bar{z}))\right) &= 0 \\
%\partial_{\bar{a}} \left(g^{i\bar{j}} \partial_i \mathrm{log}(K(z,\bar{z}))\; \partial_{\bar{j}} \mathrm{log}(K(z,\bar{z}))\right) &= 0 
%\end{align*}
%To see the first, we note
%\begin{align*}
%&\partial_a \left(g^{i\bar{j}} \partial_i \mathrm{log}(K(z,\bar{z}))\; \partial_{\bar{j}} \mathrm{log}(K(z,\bar{z}))\right) \\
%&=~ g^{i\bar{j}}\left(\partial_i \partial_a \mathrm{log}(K)\partial_{\bar{j}}\mathrm{log}(K) + \partial_i \mathrm{log}(K)\partial_{\bar{j}}\partial_a \mathrm{log}(K)\right)\\
%&= -\delta^{\bar{j}}_a  \partial_{\bar{j}} K + g^{i\bar{j}}\partial_i K g_{\bar{j} a}= -\partial_a K + \partial_a K  = 0
%\end{align*}
% \fr{I still don't understand this. Maybe we can discuss this on zoom tomorrow?}
% \BS{added details}
% \fr{$g_{ia}=0$ for Kahler metrics, so the first term is zero. Why is $g^{i\bar{j}}\partial_i K g_{\bar{j} a}=\partial_a K=0$?}
%where the second equality follows from the fact that $K(z,\bar{z})$ is a function only of $z- \bar{z}$ in this case, and hence
%\[
%\partial_i(\mathrm{log}(K)) = - \partial_{\bar{i}}(\mathrm{log}(K))
%\]
%together with the definition of the metric from the K\"ahler potential.
%The second vanishing follows from an identical argument.

To prove the second claim, we note that under the action of $\Phi$:
\begin{align*}
\nabla_a(\nabla^a F) &= g^{i\bar{j}} \partial_i \partial_{\bar{j}} \mathrm{log} \left[\frac{Z_{\tau}(\mathcal{E})}{\sqrt{K(z,\bar{z})}}\right] \\
&= g^{i\bar{j}}\partial_i \partial_{\bar{j}}\mathrm{log}\left[\frac{1}{\sqrt{K(z,\bar{z})}}\right] \\&= -\frac{1}{2} g^{i\bar{j}} \partial_i \partial_{\bar{j}} \mathrm{log}[K(z,\bar{z})] = -\frac{1}{2}g^{i\bar{j}} g_{i\bar{j}} 
\end{align*}
which must also be constant.

We give the simplest example of a single $T^2$ with $\tau=\tau_1+i\tau_2$, following~\cite{Ahmed:2023cnw}. In this case, we find from~\eqref{eq:DualVector}:
\begin{align*}
    c=\frac{2\pi}{\tau_2} (k_1\tau_2, k_2-k_1\tau_1)\, ~\Rightarrow~e_{\vec{k}}=\frac{4\pi^2}{\tau_2}|k_2-k_1\tau|^2.
\end{align*}
The (normalized) central charge of a BPS object with charges $\vec{Q}=(p,q)$ is given by
\begin{align}
    Z(\vec{Q}) = e^{K/2}(\vec{Q}\cdot\chi\cdot\Pi)=   
    \frac{1}{\sqrt{\tau_2}} (q-p\tau)\,,
\end{align}
where the symplectic scalar product is implemented via
\begin{align}
    \chi=\begin{pmatrix}0&-1\\1&0\end{pmatrix}\,,
\end{align}
and $K=-\log\tau_2$ is the K\"ahler potential for the Weil-Petersson metric in the Siegel upper half space. 

We thus observe that the normalized eigenvalues $e_{\vec{k}}$ of the Laplacian on $T^2$ are related to the eigenfunctions $F=|Z|$ of the Laplacian on the moduli space of $T^2$ via
\begin{align}
\label{eq:EigenvaluesERigenfunctions}
    e_{\vec{k}=\vec{Q}}=4\pi^2 |Z|^2
\end{align}
In~\cite{Ahmed:2023cnw}, it was shown that in this case, certain degeneration loci of eigenvalues of the scalar spacetime Laplacian (the level crossings first observed in~\cite{Ashmore:2021qdf} for the scalar Laplacian on the Quintic) are related to complex multiplication, and hence, by the result of Moore~\cite{Moore:1998pn}, to attractor points. The condition that $(\nabla F)^2$ and $\Delta F$ are constant are intimately related to the attractor mechanism, which in turn establishes a relation between the eigenvalues and the eigenfunctions of the moduli space Laplacian. This reinforces the observation of~\cite{Ahmed:2023cnw} which relates Laplacian eigenmodes and attractor points.
%where seemingly non-BPS objects (spacetime Laplacian eigenvalues) were related to BPS objects (attractor points). One of the explanations that was put forward and that seems to be correct from this analysis is that while the higher Laplacian eigenmodes are non-BPS, the crossing loci which lie at attractor points in the complex structure moduli space correspond to masses of the BPS states. 
It is a very interesting question whether this generalizes, i.e., whether the following questions based on observations in~\cite{Etheredge:2023zjk} and \cite{Ahmed:2023cnw}, respectively, are true more generally: Let $e_{\vec{k}}$ be an eigenvalue of the Laplacian on a CY $X$.\\
\textbf{Question 1:}\footnote{M.~Etheredge raised this in private communication.} Then, $e_{\vec{k}}=e_{\vec{k}}(z)$ is a function on the moduli space $\mathcal{M}$ of $X$ with coordinates $z$. Is $e_{\vec{k}}(z)$ an eigenfunction of the Laplacian on $K$?\\
\textbf{Question 2:} Then,  $e_{\vec{k}}$ is in general not BPS. Which codimension $d>0$ loci in $\mathcal{M}$ with $e_{\vec{k}}=e_{\vec{k'}}$ for $\vec{k}\neq\vec{k}'$ are attractor loci of BPS states? An do they satisfy $e_{\vec{k}}\propto|Z|^2$ as in~\eqref{eq:EigenvaluesERigenfunctions}?

Preliminary computations for the case $T^4$ suggest that some modification of Question~$1$ must be made. Indeed, for the vector:
\begin{align*}
\vec{k} = (k_1,k_2,k_3,k_4), \quad v_k = \vec{k}^T\cdot\text{Cof}(\Pi)^T\text{Cof}(\Pi)\cdot\vec{k}   
\end{align*}
the rescaled eigenvalue 
\begin{align}
\label{eq:EigenvaluesT4}
\left(\frac{1}{\mathrm{det}(\Pi)}\right)^nv_k
\end{align}
is not an eigenfunction of the Laplacian on the Siegel upper half space with metric~(\ref{eqn:metric}) for any rational $n$.

Nevertheless, equation~\eqref{eq:EigenvaluesT4} seems to hold for $n = 1$ when restricted to the diagonal sub-locus of the Siegel space where $T^4=T^2_{\tau_1}\times T^2_{\tau_2}$, i.e., to diagonal matrices in~(\ref{eqn:siegel}). In addition, certain eigenvalue functions $e_{\vec{k}}$ coincide with the central charges of certain BPS states; more precisely equation~(\ref{eq:EigenvaluesERigenfunctions}) holds along this loci on $T^4$ with 
\[
Q = (q_1, q_2, 0,0), \quad \vec{k} = (0,0, q_2 ,q_1).
\]More generally, we have observed that the eigenvalue functions $e_{\vec{k}}$ coincide with central charge functions near attractor points, and it would be desirable to understand if they correspond to masses of various KK/winding modes, and to verify the existence of additional BPS states corresponding to these functions at crossing points.

In our analysis, we relied on the existence of a large symmetry group $\mathrm{Sp}(n)$ which acts naturally on the RR-charges. In particular, this acts naturally in the Siegel upper half space, which simplifies the study of attractor flows associated with different BPS defects. While we have no proof that the Coxeter group captures the full symmetry group of birational automorphisms on a generic Calabi-Yau threefold $X$, we are not aware of a larger symmetry and expect it will be very useful in studies of the uniqueness of geodesics. We hope to return to this in the future.

\section{The one-parameter family of Quintics}
As a second example, let us discuss the one-parameter family of Quintics. This is perhaps the simplest example for a bona fide CY threefold and has served as a prime example in the context of geodesics~\cite{Ashmore:2021qdf,Blumenhagen:2018nts}, attractors~\cite{Denef:2000nb,Denef:2001xn,Ahmed:2023cnw}, and marked moduli spaces~\cite{Raman:2024fcv}. Its moduli space is the first one that had been studied in the development of mirror symmetry in the seminal paper~\cite{Candelas:1990rm}.

The Quintic is defined as a section of the anti-canonical bundle in $\mathbbm{P}^4$, which is given by the zero locus of a homogeneous polynomial $p=0$ of degree 5 in the homogeneous coordinates $z_i$ of $\mathbbm{P}^4$.\footnote{Note that now $z_i$ are spacetime coordinates while $\psi$ is the coordinate on the marked moduli space and $\psi^5$ is the coordinate on the moduli space, see below.} Restricting to a symmetric locus leaves only one independent complex structure modulus $\psi$,
\begin{align}
    p=\sum_{i=0}^4z_i^5 -5\psi\prod_{i=0}^4z_i=0\,.
\end{align}
The shooting problem of finding geodesics that connects two points $\psi_0$ and $\psi_1$ was solved numerically in~\cite{Ashmore:2021qdf}. The geodesics were also studied in~\cite{Raman:2024fcv,Blumenhagen:2018nts}, while the attractor flow was worked out numerically in~\cite{Denef:2001xn}.

The moduli space has a Gepner point at $\psi=0$, a conifold point at $\psi=1$, and a large complex structure (LCS) point at $|\psi|\to\infty$. The fact that $\psi^5$ is the coordinate of the moduli space can be seen from the fact that $\psi\to e^{2\pi i/5}\psi$ can be absorbed in a coordinate redefinition. Alternatively, by solving the hypergeometric Picard-Fuchs system to obtain the periods, one finds that the fundamental period is a hypergeometric function involving $\psi^5$ instead of $\psi$. In going from the moduli space defined in a wedge $0\leq\arg(\psi)<2\pi/5$, say, to the marked moduli space parameterized by $\psi$, one has to take monodromies around the LCS point into account.

It was argued in~\cite{Raman:2024fcv} that precisely in this case geodesics from the LG point to a point $\psi>1$ on the real axis are unique in the marked moduli space, since they can be distinguished by the monodromy frame, similar to how passing from the fundamental domain of the modulus parameter $\tau$ of a torus to the entire Siegel upper half space distinguishes points by their SL$(2,\mathbbm{Z})$ charge.

\begin{figure}
    \centering
    \includegraphics[width=.8\linewidth]{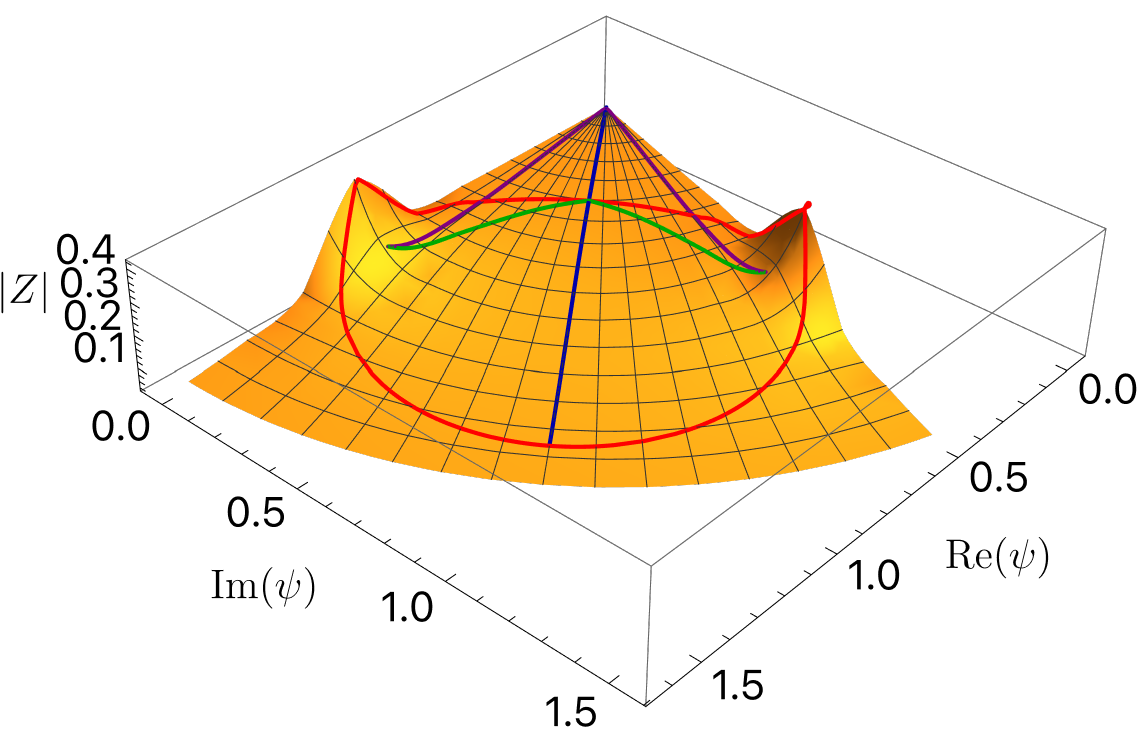}
    \caption{Split attractor flows. The red line delineates the wall of marginal stability. The blue line is the attractor flow (i.e., a geodesic) for a particle of charge $\Gamma$ from the LG point to the attractor point. The green lines are the attractor flows of $\Gamma_1$ and $\Gamma_2$ towords their attractor points (which is the conifold point). The purple lines are geodesics from the LG point to the conifold points. All flows are plotted on top of the Weil-Petersson moduli space metric.}
    \label{fig:AttractorFlow}
\end{figure}

Flows from the Gepner point towards the LCS point have also been studied in the context of attractors in~\cite{Denef:2000nb,Denef:2001xn}. If $Z\neq0$, the flow follows gradient descent in the potential $|Z|$, while flows towards attractor points with $|Z|=0$ are more special, as discussed above: Moore points out~\cite{Moore:1998pn} that attractor flows of this type ($Z=0$ in the interior of the moduli space) break down and the charge does not support a BPS state. However, starting at $\psi=0$, there is a BPS state $\Gamma$, which crosses a wall of marginal stability along its flow to the fixed point with $Z=0$. At the wall, it decays into two lighter BPS particles $\Gamma\to\Gamma_1+\Gamma_2$. Instead of reaching the end point $Z=0$, one particle gets stuck at the wall while the other continues along an attractor flow with a different fixed point $Z(\Gamma_i)\neq0$, see Figure~\ref{fig:AttractorFlow} for an example. 
In this example (also studied in~\cite{Denef:2000nb}), we consider a particle with charge
\begin{align}
    \Gamma=(Q_6,Q_4,Q_2,Q_0)=(2,1,3,-5)\,,
\end{align}
whose attractor flow starts at the LG point $\psi=0$ and ends at $Z=0$, $\psi\approx 1.46$. Around $\psi\approx0.51$, it crosses a wall of marginal stability and the attractor flow splits into two,
\begin{align}
\begin{split}
\Gamma&\to\Gamma_1+\Gamma_2\\
(2,1,3,-5)&\to (1, 0, 0, 0) + (1, 1, 3, -5)\,,
\end{split}
\end{align}
which end at two copies of the conifold point. At the wall of marginal stability, we have 
\begin{align}
    Z(\Gamma)=Z(\Gamma_1)+Z(\Gamma_2)\,,
\end{align}
and $\Gamma$ is strictly heavier than $\Gamma_1+\Gamma_2$ in the region enclosed by the wall. The attractor flows for $\Gamma_1$ and $\Gamma_2$ correspond to two inequivalent flow lines from the Gepner point to the conifold point. In particular, from the split point at the wall around $\psi\approx0.51$ to the conifold point $\psi=1\equiv e^{2\pi i/5}$, there are two inequivalent geodesics corresponding to the attractor flows of $\Gamma_{1,2}$. In the marked moduli space, the conifold points are again distinguished by their D-brane charges, or equivalently, by the monodromy action. 

We should point out that the full split flow from the LG to the conifold point is \textit{not} geodesic, but comprised of two geodesics: one corresponding to the attractor flow up until the wall of marginal stability, and one from the wall to the attractor point. The geodesics from the LG point (with $\psi\approx \varepsilon_1 e^{i (2\pi/10 \pm\varepsilon_2)}$) to the conifold points ($\psi=1$ or $\psi=e^{2\pi i/5}$) are depicted in purple.

In the picture of split flows, a necessary condition for the uniqueness of geodesics is the absence of loop junctions (in the interior of the moduli space). If a loop junction existed, the geodesic from the split point to any other point on the loop would be non-unique. Typically, such string junctions are not stable: they either collapse     or correspond to normal junctions under Hanany-Witten moves, i.e., are distinguished by monodromy.

\section{Conclusions}
A recurrent theme in the past decade has been the rediscovery of general properties of high-dimensional, and highly supersymmetric string compactifications based on general principles of quantum gravity, motivated primarily by the swampland program. This stands in contrast with, and is complemented by progress from the early days of string theory, where much effort was focused on explicit constructions of various corners of the string landscape, leading to empirical derivations of deep physical principles such as dualities. One of the most tractable features of supersymmetric string compactifications is the geometry and topology of the configuration space of both the massless and massive fields; nevertheless, much remains to be explored even in the simplest models.

In this paper, we point out and make explicit the connections among various aspects of moduli space geometry: contractibility, geodesics, attractor flows, and Laplacians. We discuss concrete examples primarily for supersymmetric string compactifications on abelian varieties, as well as on Calabi-Yau threefolds. In the former class of examples, we describe the connection between Laplacian eigenvalue crossings and attractor points, and how both mechanisms seem to give complementary descriptions for dynamical realizations of BPS states. In the latter class, we elucidate the physical interpretation of geodesics as attractor flows, and we demonstrate first steps towards using this to prove uniqueness of geodesics in $5$d $\mathcal{N}=1$ supergravity theories. 

There are many remaining questions to be explored in detail that we leave for future work. For one, it would be desirable to study further the metric and geodesic properties of $4$d $\mathcal{N} = 2$ compactifications and their physical interpretations as attractor flows. The moduli space of abelian varieties is a symmetric space, while the moduli space for K3-surfaces admits a period mapping to a period domain which is symmetric. In general, and in particular for Calabi-Yau threefolds, there are less symmetry groups to exploit; nevertheless, the period mapping seems critical for physical applications and this was used to define and exploit the Hodge metric for contractibility~\cite{Raman:2024fcv} recently. For another, the precise details and degeneracies of BPS states and their algebras~\cite{Harvey:1996gc,Harvey:1995fq}, and their dynamical realizations via flows even for maximal supergravity theories remain to be explored, together with a number of emerging complementary descriptions via attractor points, Hodge loci~\cite{Grimm:2024fip}, and Laplacian eigenvalue crossings.

\section*{Acknowledgments}
We thank Muldrow Etheredge, Sergei Gukov, Eric Sharpe, and Damian van de Heisteeg for helpful discussions. The work of FR is supported by the NSF grants PHY-2210333 and PHY-2019786 (The NSF AI Institute for Artificial Intelligence and Fundamental Interactions), as well as startup funding from Northeastern University. 
\bibliography{bibliography}

\begin{thebibliography}{23}%
\makeatletter
\providecommand \@ifxundefined [1]{%
 \@ifx{#1\undefined}
}%
\providecommand \@ifnum [1]{%
 \ifnum #1\expandafter \@firstoftwo
 \else \expandafter \@secondoftwo
 \fi
}%
\providecommand \@ifx [1]{%
 \ifx #1\expandafter \@firstoftwo
 \else \expandafter \@secondoftwo
 \fi
}%
\providecommand \natexlab [1]{#1}%
\providecommand \enquote  [1]{``#1''}%
\providecommand \bibnamefont  [1]{#1}%
\providecommand \bibfnamefont [1]{#1}%
\providecommand \citenamefont [1]{#1}%
\providecommand \href@noop [0]{\@secondoftwo}%
\providecommand \href [0]{\begingroup \@sanitize@url \@href}%
\providecommand \@href[1]{\@@startlink{#1}\@@href}%
\providecommand \@@href[1]{\endgroup#1\@@endlink}%
\providecommand \@sanitize@url [0]{\catcode `\\12\catcode `\$12\catcode
  `\&12\catcode `\#12\catcode `\^12\catcode `\_12\catcode `\%12\relax}%
\providecommand \@@startlink[1]{}%
\providecommand \@@endlink[0]{}%
\providecommand \url  [0]{\begingroup\@sanitize@url \@url }%
\providecommand \@url [1]{\endgroup\@href {#1}{\urlprefix }}%
\providecommand \urlprefix  [0]{URL }%
\providecommand \Eprint [0]{\href }%
\providecommand \doibase [0]{http://dx.doi.org/}%
\providecommand \selectlanguage [0]{\@gobble}%
\providecommand \bibinfo  [0]{\@secondoftwo}%
\providecommand \bibfield  [0]{\@secondoftwo}%
\providecommand \translation [1]{[#1]}%
\providecommand \BibitemOpen [0]{}%
\providecommand \bibitemStop [0]{}%
\providecommand \bibitemNoStop [0]{.\EOS\space}%
\providecommand \EOS [0]{\spacefactor3000\relax}%
\providecommand \BibitemShut  [1]{\csname bibitem#1\endcsname}%
\let\auto@bib@innerbib\@empty
%</preamble>
\bibitem [{\citenamefont {Ahmed}\ and\ \citenamefont
  {Ruehle}(2023)}]{Ahmed:2023cnw}%
  \BibitemOpen
  \bibfield  {author} {\bibinfo {author} {\bibfnamefont {H.}~\bibnamefont
  {Ahmed}}\ and\ \bibinfo {author} {\bibfnamefont {F.}~\bibnamefont {Ruehle}},\
  }\href {\doibase 10.1007/JHEP06(2023)164} {\bibfield  {journal} {\bibinfo
  {journal} {JHEP}\ }\textbf {\bibinfo {volume} {06}},\ \bibinfo {pages} {164}
  (\bibinfo {year} {2023})},\ \Eprint {http://arxiv.org/abs/2304.00027}
  {arXiv:2304.00027 [hep-th]} \BibitemShut {NoStop}%
\bibitem [{\citenamefont {Etheredge}\ and\ \citenamefont
  {Heidenreich}(2023)}]{Etheredge:2023zjk}%
  \BibitemOpen
  \bibfield  {author} {\bibinfo {author} {\bibfnamefont {M.}~\bibnamefont
  {Etheredge}}\ and\ \bibinfo {author} {\bibfnamefont {B.}~\bibnamefont
  {Heidenreich}},\ }\href@noop {} {\  (\bibinfo {year} {2023})},\ \Eprint
  {http://arxiv.org/abs/2311.18693} {arXiv:2311.18693 [hep-th]} \BibitemShut
  {NoStop}%
\bibitem [{\citenamefont {Raman}\ and\ \citenamefont
  {Vafa}(2024)}]{Raman:2024fcv}%
  \BibitemOpen
  \bibfield  {author} {\bibinfo {author} {\bibfnamefont {S.}~\bibnamefont
  {Raman}}\ and\ \bibinfo {author} {\bibfnamefont {C.}~\bibnamefont {Vafa}},\
  }\href@noop {} {\  (\bibinfo {year} {2024})},\ \Eprint
  {http://arxiv.org/abs/2405.11611} {arXiv:2405.11611 [hep-th]} \BibitemShut
  {NoStop}%
\bibitem [{\citenamefont {Ferrara}\ \emph {et~al.}(1995)\citenamefont
  {Ferrara}, \citenamefont {Kallosh},\ and\ \citenamefont
  {Strominger}}]{Ferrara:1995ih}%
  \BibitemOpen
  \bibfield  {author} {\bibinfo {author} {\bibfnamefont {S.}~\bibnamefont
  {Ferrara}}, \bibinfo {author} {\bibfnamefont {R.}~\bibnamefont {Kallosh}}, \
  and\ \bibinfo {author} {\bibfnamefont {A.}~\bibnamefont {Strominger}},\
  }\href {\doibase 10.1103/PhysRevD.52.R5412} {\bibfield  {journal} {\bibinfo
  {journal} {Phys. Rev. D}\ }\textbf {\bibinfo {volume} {52}},\ \bibinfo
  {pages} {R5412} (\bibinfo {year} {1995})},\ \Eprint
  {http://arxiv.org/abs/hep-th/9508072} {arXiv:hep-th/9508072} \BibitemShut
  {NoStop}%
\bibitem [{\citenamefont {Ferrara}\ \emph {et~al.}(1997)\citenamefont
  {Ferrara}, \citenamefont {Gibbons},\ and\ \citenamefont
  {Kallosh}}]{Ferrara:1997tw}%
  \BibitemOpen
  \bibfield  {author} {\bibinfo {author} {\bibfnamefont {S.}~\bibnamefont
  {Ferrara}}, \bibinfo {author} {\bibfnamefont {G.~W.}\ \bibnamefont
  {Gibbons}}, \ and\ \bibinfo {author} {\bibfnamefont {R.}~\bibnamefont
  {Kallosh}},\ }\href {\doibase 10.1016/S0550-3213(97)00324-6} {\bibfield
  {journal} {\bibinfo  {journal} {Nucl. Phys. B}\ }\textbf {\bibinfo {volume}
  {500}},\ \bibinfo {pages} {75} (\bibinfo {year} {1997})},\ \Eprint
  {http://arxiv.org/abs/hep-th/9702103} {arXiv:hep-th/9702103} \BibitemShut
  {NoStop}%
\bibitem [{\citenamefont {Moore}(1998)}]{Moore:1998pn}%
  \BibitemOpen
  \bibfield  {author} {\bibinfo {author} {\bibfnamefont {G.~W.}\ \bibnamefont
  {Moore}},\ }\href@noop {} {\  (\bibinfo {year} {1998})},\ \Eprint
  {http://arxiv.org/abs/hep-th/9807087} {arXiv:hep-th/9807087} \BibitemShut
  {NoStop}%
\bibitem [{\citenamefont {Denef}(2000)}]{Denef:2000nb}%
  \BibitemOpen
  \bibfield  {author} {\bibinfo {author} {\bibfnamefont {F.}~\bibnamefont
  {Denef}},\ }\href {\doibase 10.1088/1126-6708/2000/08/050} {\bibfield
  {journal} {\bibinfo  {journal} {JHEP}\ }\textbf {\bibinfo {volume} {08}},\
  \bibinfo {pages} {050} (\bibinfo {year} {2000})},\ \Eprint
  {http://arxiv.org/abs/hep-th/0005049} {arXiv:hep-th/0005049} \BibitemShut
  {NoStop}%
\bibitem [{\citenamefont {Denef}\ \emph {et~al.}(2001)\citenamefont {Denef},
  \citenamefont {Greene},\ and\ \citenamefont {Raugas}}]{Denef:2001xn}%
  \BibitemOpen
  \bibfield  {author} {\bibinfo {author} {\bibfnamefont {F.}~\bibnamefont
  {Denef}}, \bibinfo {author} {\bibfnamefont {B.~R.}\ \bibnamefont {Greene}}, \
  and\ \bibinfo {author} {\bibfnamefont {M.}~\bibnamefont {Raugas}},\ }\href
  {\doibase 10.1088/1126-6708/2001/05/012} {\bibfield  {journal} {\bibinfo
  {journal} {JHEP}\ }\textbf {\bibinfo {volume} {05}},\ \bibinfo {pages} {012}
  (\bibinfo {year} {2001})},\ \Eprint {http://arxiv.org/abs/hep-th/0101135}
  {arXiv:hep-th/0101135} \BibitemShut {NoStop}%
\bibitem [{\citenamefont {Lam}\ and\ \citenamefont
  {Tripathy}(2024)}]{Lam:2020qge}%
  \BibitemOpen
  \bibfield  {author} {\bibinfo {author} {\bibfnamefont {Y.~H.~J.}\
  \bibnamefont {Lam}}\ and\ \bibinfo {author} {\bibfnamefont {A.}~\bibnamefont
  {Tripathy}},\ }\href {\doibase 10.1112/S0010437X24007036} {\bibfield
  {journal} {\bibinfo  {journal} {Compos. Math.}\ }\textbf {\bibinfo {volume}
  {160}},\ \bibinfo {pages} {1073} (\bibinfo {year} {2024})},\ \Eprint
  {http://arxiv.org/abs/2009.12650} {arXiv:2009.12650 [math.NT]} \BibitemShut
  {NoStop}%
\bibitem [{\citenamefont {Brodie}\ \emph {et~al.}(2022)\citenamefont {Brodie},
  \citenamefont {Constantin}, \citenamefont {Lukas},\ and\ \citenamefont
  {Ruehle}}]{Brodie:2021nit}%
  \BibitemOpen
  \bibfield  {author} {\bibinfo {author} {\bibfnamefont {C.~R.}\ \bibnamefont
  {Brodie}}, \bibinfo {author} {\bibfnamefont {A.}~\bibnamefont {Constantin}},
  \bibinfo {author} {\bibfnamefont {A.}~\bibnamefont {Lukas}}, \ and\ \bibinfo
  {author} {\bibfnamefont {F.}~\bibnamefont {Ruehle}},\ }\href {\doibase
  10.1007/JHEP03(2022)024} {\bibfield  {journal} {\bibinfo  {journal} {JHEP}\
  }\textbf {\bibinfo {volume} {03}},\ \bibinfo {pages} {024} (\bibinfo {year}
  {2022})},\ \Eprint {http://arxiv.org/abs/2108.10323} {arXiv:2108.10323
  [hep-th]} \BibitemShut {NoStop}%
\bibitem [{\citenamefont {Lukas}\ and\ \citenamefont
  {Ruehle}(2023)}]{Lukas:2022crp}%
  \BibitemOpen
  \bibfield  {author} {\bibinfo {author} {\bibfnamefont {A.}~\bibnamefont
  {Lukas}}\ and\ \bibinfo {author} {\bibfnamefont {F.}~\bibnamefont {Ruehle}},\
  }\href {\doibase 10.1007/JHEP02(2023)175} {\bibfield  {journal} {\bibinfo
  {journal} {JHEP}\ }\textbf {\bibinfo {volume} {02}},\ \bibinfo {pages} {175}
  (\bibinfo {year} {2023})},\ \Eprint {http://arxiv.org/abs/2210.09369}
  {arXiv:2210.09369 [hep-th]} \BibitemShut {NoStop}%
\bibitem [{\citenamefont {Loi}\ and\ \citenamefont
  {Mossa}(2015)}]{Loi:2015aaa}%
  \BibitemOpen
  \bibfield  {author} {\bibinfo {author} {\bibfnamefont {A.}~\bibnamefont
  {Loi}}\ and\ \bibinfo {author} {\bibfnamefont {R.}~\bibnamefont {Mossa}},\
  }\href {\doibase 10.1007/s10711-015-0085-5} {\bibfield  {journal} {\bibinfo
  {journal} {Geometriae Dedicata}\ }\textbf {\bibinfo {volume} {179}},\
  \bibinfo {pages} {377–383} (\bibinfo {year} {2015})}\BibitemShut {NoStop}%
\bibitem [{\citenamefont {Gomis}\ \emph {et~al.}(2016)\citenamefont {Gomis},
  \citenamefont {Hsin}, \citenamefont {Komargodski}, \citenamefont {Schwimmer},
  \citenamefont {Seiberg},\ and\ \citenamefont {Theisen}}]{Gomis:2015yaa}%
  \BibitemOpen
  \bibfield  {author} {\bibinfo {author} {\bibfnamefont {J.}~\bibnamefont
  {Gomis}}, \bibinfo {author} {\bibfnamefont {P.-S.}\ \bibnamefont {Hsin}},
  \bibinfo {author} {\bibfnamefont {Z.}~\bibnamefont {Komargodski}}, \bibinfo
  {author} {\bibfnamefont {A.}~\bibnamefont {Schwimmer}}, \bibinfo {author}
  {\bibfnamefont {N.}~\bibnamefont {Seiberg}}, \ and\ \bibinfo {author}
  {\bibfnamefont {S.}~\bibnamefont {Theisen}},\ }\href {\doibase
  10.1007/JHEP03(2016)022} {\bibfield  {journal} {\bibinfo  {journal} {JHEP}\
  }\textbf {\bibinfo {volume} {03}},\ \bibinfo {pages} {022} (\bibinfo {year}
  {2016})},\ \Eprint {http://arxiv.org/abs/1509.08511} {arXiv:1509.08511
  [hep-th]} \BibitemShut {NoStop}%
\bibitem [{\citenamefont {Donagi}\ \emph {et~al.}(2022)\citenamefont {Donagi},
  \citenamefont {Macerato},\ and\ \citenamefont {Sharpe}}]{Donagi:2017mhd}%
  \BibitemOpen
  \bibfield  {author} {\bibinfo {author} {\bibfnamefont {R.}~\bibnamefont
  {Donagi}}, \bibinfo {author} {\bibfnamefont {M.}~\bibnamefont {Macerato}}, \
  and\ \bibinfo {author} {\bibfnamefont {E.}~\bibnamefont {Sharpe}},\ }\href
  {\doibase 10.4310/AJM.2022.v26.n4.a4} {\bibfield  {journal} {\bibinfo
  {journal} {Asian J. Math.}\ }\textbf {\bibinfo {volume} {26}},\ \bibinfo
  {pages} {585} (\bibinfo {year} {2022})},\ \Eprint
  {http://arxiv.org/abs/1707.05322} {arXiv:1707.05322 [math.AG]} \BibitemShut
  {NoStop}%
\bibitem [{\citenamefont {Brodie}\ \emph {et~al.}(2021)\citenamefont {Brodie},
  \citenamefont {Constantin}, \citenamefont {Lukas},\ and\ \citenamefont
  {Ruehle}}]{Brodie:2021ain}%
  \BibitemOpen
  \bibfield  {author} {\bibinfo {author} {\bibfnamefont {C.~R.}\ \bibnamefont
  {Brodie}}, \bibinfo {author} {\bibfnamefont {A.}~\bibnamefont {Constantin}},
  \bibinfo {author} {\bibfnamefont {A.}~\bibnamefont {Lukas}}, \ and\ \bibinfo
  {author} {\bibfnamefont {F.}~\bibnamefont {Ruehle}},\ }\href {\doibase
  10.1103/PhysRevD.104.046008} {\bibfield  {journal} {\bibinfo  {journal}
  {Phys. Rev. D}\ }\textbf {\bibinfo {volume} {104}},\ \bibinfo {pages}
  {046008} (\bibinfo {year} {2021})},\ \Eprint
  {http://arxiv.org/abs/2104.03325} {arXiv:2104.03325 [hep-th]} \BibitemShut
  {NoStop}%
\bibitem [{\citenamefont {Etheredge}\ \emph {et~al.}(2022)\citenamefont
  {Etheredge}, \citenamefont {Heidenreich}, \citenamefont {Kaya}, \citenamefont
  {Qiu},\ and\ \citenamefont {Rudelius}}]{Etheredge_2022}%
  \BibitemOpen
  \bibfield  {author} {\bibinfo {author} {\bibfnamefont {M.}~\bibnamefont
  {Etheredge}}, \bibinfo {author} {\bibfnamefont {B.}~\bibnamefont
  {Heidenreich}}, \bibinfo {author} {\bibfnamefont {S.}~\bibnamefont {Kaya}},
  \bibinfo {author} {\bibfnamefont {Y.}~\bibnamefont {Qiu}}, \ and\ \bibinfo
  {author} {\bibfnamefont {T.}~\bibnamefont {Rudelius}},\ }\href {\doibase
  10.1007/jhep12(2022)114} {\bibfield  {journal} {\bibinfo  {journal} {Journal
  of High Energy Physics}\ }\textbf {\bibinfo {volume} {2022}} (\bibinfo {year}
  {2022}),\ 10.1007/jhep12(2022)114}\BibitemShut {NoStop}%
\bibitem [{\citenamefont {Etheredge}(2024)}]{Etheredge:2023usk}%
  \BibitemOpen
  \bibfield  {author} {\bibinfo {author} {\bibfnamefont {M.}~\bibnamefont
  {Etheredge}},\ }\href {\doibase 10.1007/JHEP01(2024)122} {\bibfield
  {journal} {\bibinfo  {journal} {JHEP}\ }\textbf {\bibinfo {volume} {01}},\
  \bibinfo {pages} {122} (\bibinfo {year} {2024})},\ \Eprint
  {http://arxiv.org/abs/2308.01331} {arXiv:2308.01331 [hep-th]} \BibitemShut
  {NoStop}%
\bibitem [{\citenamefont {Ashmore}\ and\ \citenamefont
  {Ruehle}(2021)}]{Ashmore:2021qdf}%
  \BibitemOpen
  \bibfield  {author} {\bibinfo {author} {\bibfnamefont {A.}~\bibnamefont
  {Ashmore}}\ and\ \bibinfo {author} {\bibfnamefont {F.}~\bibnamefont
  {Ruehle}},\ }\href {\doibase 10.1103/PhysRevD.103.106028} {\bibfield
  {journal} {\bibinfo  {journal} {Phys. Rev. D}\ }\textbf {\bibinfo {volume}
  {103}},\ \bibinfo {pages} {106028} (\bibinfo {year} {2021})},\ \Eprint
  {http://arxiv.org/abs/2103.07472} {arXiv:2103.07472 [hep-th]} \BibitemShut
  {NoStop}%
\bibitem [{\citenamefont {Blumenhagen}\ \emph {et~al.}(2018)\citenamefont
  {Blumenhagen}, \citenamefont {Kl\"awer}, \citenamefont {Schlechter},\ and\
  \citenamefont {Wolf}}]{Blumenhagen:2018nts}%
  \BibitemOpen
  \bibfield  {author} {\bibinfo {author} {\bibfnamefont {R.}~\bibnamefont
  {Blumenhagen}}, \bibinfo {author} {\bibfnamefont {D.}~\bibnamefont
  {Kl\"awer}}, \bibinfo {author} {\bibfnamefont {L.}~\bibnamefont
  {Schlechter}}, \ and\ \bibinfo {author} {\bibfnamefont {F.}~\bibnamefont
  {Wolf}},\ }\href {\doibase 10.1007/JHEP06(2018)052} {\bibfield  {journal}
  {\bibinfo  {journal} {JHEP}\ }\textbf {\bibinfo {volume} {06}},\ \bibinfo
  {pages} {052} (\bibinfo {year} {2018})},\ \Eprint
  {http://arxiv.org/abs/1803.04989} {arXiv:1803.04989 [hep-th]} \BibitemShut
  {NoStop}%
\bibitem [{\citenamefont {Candelas}\ \emph {et~al.}(1991)\citenamefont
  {Candelas}, \citenamefont {De~La~Ossa}, \citenamefont {Green},\ and\
  \citenamefont {Parkes}}]{Candelas:1990rm}%
  \BibitemOpen
  \bibfield  {author} {\bibinfo {author} {\bibfnamefont {P.}~\bibnamefont
  {Candelas}}, \bibinfo {author} {\bibfnamefont {X.~C.}\ \bibnamefont
  {De~La~Ossa}}, \bibinfo {author} {\bibfnamefont {P.~S.}\ \bibnamefont
  {Green}}, \ and\ \bibinfo {author} {\bibfnamefont {L.}~\bibnamefont
  {Parkes}},\ }\href {\doibase 10.1016/0550-3213(91)90292-6} {\bibfield
  {journal} {\bibinfo  {journal} {Nucl. Phys. B}\ }\textbf {\bibinfo {volume}
  {359}},\ \bibinfo {pages} {21} (\bibinfo {year} {1991})}\BibitemShut
  {NoStop}%
\bibitem [{\citenamefont {Harvey}\ and\ \citenamefont
  {Moore}(1998)}]{Harvey:1996gc}%
  \BibitemOpen
  \bibfield  {author} {\bibinfo {author} {\bibfnamefont {J.~A.}\ \bibnamefont
  {Harvey}}\ and\ \bibinfo {author} {\bibfnamefont {G.~W.}\ \bibnamefont
  {Moore}},\ }\href {\doibase 10.1007/s002200050461} {\bibfield  {journal}
  {\bibinfo  {journal} {Commun. Math. Phys.}\ }\textbf {\bibinfo {volume}
  {197}},\ \bibinfo {pages} {489} (\bibinfo {year} {1998})},\ \Eprint
  {http://arxiv.org/abs/hep-th/9609017} {arXiv:hep-th/9609017} \BibitemShut
  {NoStop}%
\bibitem [{\citenamefont {Harvey}\ and\ \citenamefont
  {Moore}(1996)}]{Harvey:1995fq}%
  \BibitemOpen
  \bibfield  {author} {\bibinfo {author} {\bibfnamefont {J.~A.}\ \bibnamefont
  {Harvey}}\ and\ \bibinfo {author} {\bibfnamefont {G.~W.}\ \bibnamefont
  {Moore}},\ }\href {\doibase 10.1016/0550-3213(95)00605-2} {\bibfield
  {journal} {\bibinfo  {journal} {Nucl. Phys. B}\ }\textbf {\bibinfo {volume}
  {463}},\ \bibinfo {pages} {315} (\bibinfo {year} {1996})},\ \Eprint
  {http://arxiv.org/abs/hep-th/9510182} {arXiv:hep-th/9510182} \BibitemShut
  {NoStop}%
\bibitem [{\citenamefont {Grimm}\ and\ \citenamefont {van~de
  Heisteeg}(2024)}]{Grimm:2024fip}%
  \BibitemOpen
  \bibfield  {author} {\bibinfo {author} {\bibfnamefont {T.~W.}\ \bibnamefont
  {Grimm}}\ and\ \bibinfo {author} {\bibfnamefont {D.}~\bibnamefont {van~de
  Heisteeg}},\ }\href@noop {} {\  (\bibinfo {year} {2024})},\ \Eprint
  {http://arxiv.org/abs/2404.12422} {arXiv:2404.12422 [hep-th]} \BibitemShut
  {NoStop}%
\end{thebibliography}%
\end{document}